\documentclass[aps,prd,twocolumn,showpacs,groupedaddress,nofootinbib]{revtex4}
\usepackage{graphicx}
\usepackage{color}
\begin{document}

\title{Critical Behavior of Ferromagnetic Ising Model on Triangular Lattice}

\author{Zhi-Huan Luo$^{a}$, Mushtaq Loan$^{b}$, Yan Liu$^{a}$\footnote{corresponding author},
and Jian-Rong Lin$^{c}$} \affiliation{ $^{a}$ Department of Applied
Physics, South China Agricultural University, Wushan Road, Guangzhou, 510642, P.R. China\\
$^{b}$ International School, Jinan University, Huangpu Road West, Guangzhou 510632, P.R. China \\
$^{c}$ Engineering College, South China Agricultural University,
Wushan Road, Guangzhou, 510642, P.R. China}

\date{\today}

\begin{abstract}
We apply a new updating algorithm scheme to investigate the critical
behavior of the two-dimensional ferromagnetic Ising model on a
triangular lattice  with nearest neighbour interactions. The
transition is examined by generating accurate data for large
lattices with $L=8,10,12,15,20,25,30,40,50$. The spin updating
algorithm we employ has the advantages of both metropolis and
single-update methods. Our study indicates that the transition to be
continuous at $T_c=3.6403(2)$. A convincing finite-size scaling
analysis of the model yield $\nu=0.9995(21)$,
$\beta/\nu=0.12400(18)$, $\gamma/\nu=1.75223(22)$,
$\gamma'/\nu=1.7555(22)$, $\alpha/\nu=0.00077(420)$ (scaling) and
$\alpha/\nu=0.0010(42)$(hyperscaling) respectively. Estimates of
present scheme yield accurate estimates for all critical exponents
than those obtained with Monte Carlo methods and show an excellent
agreement with their well-established predicted values.
\end{abstract}
\pacs{75.10.Nr, 64.60.Fr, 75.40.Cx}

\maketitle

\section{Introduction}

Ising Model on triangular lattice, as an archetypical example of a
frustrated system, was initially studied by Wannier
\cite{Wannier-1950-PR}  and Newll \cite{Newell-1950-PR}.  The
triangular Ising model has attracted much attention and because of
its remarkable properties it has a long history of investigation.
No exact solution is available in two dimensions in an arbitrary
magnetic field. Hence, simulations of the Ising model are
essential. Monte Carlo simulation methods have been widely using
techniques to update the spins of the system and to study the
Ising model on triangular lattice  to obtain numerical solutions.

A number of Monte Carlo methods based on Metropolis algorithms
\cite{Metropolis-1953-J.Chem.Phys} have been applied to the model
in the past with somewhat mixed results.  A classical Monte Carlo
using Metropolis algorithm runs into difficulties on large lattice
sizes due to critical slowing down and rapid increase in
correlation time. Because of these reasons it is difficult to
obtain meaningful results on larger lattice sizes.  A cluster
technique, which uses multi-cluster algorithm, was pioneered by
Swendsen and Wang \cite{Swendsen-1987-PRL}. Their method
demonstrated its validity and efficiency for the Potts model
successfully. Similar ideas have been pursued by Wolff
\cite{Wolff-1989-PRL} who proposed single-cluster algorithm, a
nonlocal updating technique based on multiple-cluster algorithm
but more efficient and easily applicable to achieve the meaningful
results.  Both  of these cluster algorithms are very effective in
reducing critical slowing down. But for the same system, one sweep
of single-cluster or multiple-clusters take much more time than
that of Metropolis algorithm, hence less effective for large
lattice sizes. Very little has been done since then using these
approaches on this model. Since large systems have the advantage
of reducing finite-size effects, for this reason, we are forced to
look yet again for an alternative approach.

In this study, we attempt to use a mixture of extend a
single-cluster (multiple-cluster) and Metropolis algorithms to the
Ising model in two-dimensions on triangular lattice. Applications of
combined algorithm techniques  have been extremely successful in
lattice QCD
\cite{Morningstar-1997-PRD,Morningstar-1999-PRD,Loan-2006-IJMPA,Luo-2007-MPLA,Loan08,Loan08a,
Loan08b,Loan06c}
 and have given rise to great optimism about the possibility of
obtaining results relevant to continuum physics from Monte Carlo
simulations of lattice version of the corresponding theory. As
mentioned above, our aim is to use standard Monte Carlo techniques
based on combined effect of single-cluster (multiple-cluster) and
Metropolis algorithm to update the spins of the system and see
whether useful results can be obtained. The values of the critical
exponents are well established for this model in 2-dimensions and
in order for our method to be considered successful, it must
reproduce these well-established values. For this purpose, the
critical exponents are computed by using the combined algorithm.

The rest of the paper is organised as follows:  In Sec. II we
discuss the Ising model in 2-dimensions in its lattice
formulation. Here we describe the details of simulations and the
methods used to extract the observables. We present and discuss
our results in Sec. III. Our conclusions are given in Sec. IV
\section{Model and Simulation details}

The Hamiltonian of the model is given by
\begin{equation}
H=-J\sum\limits_{<i,j>}S_iS_j, \label{eqn1}
\end{equation}
where $J$ is positive and denotes the strength of the
ferromagnetic interaction, $S_i=\pm 1$ is the Ising spin variable,
and  $<i,j>$  restricts the summation to distinct pairs of nearest
neighbors. The term on the right-hand side of Eq. \ref{eqn1} shows
that the overall energy is lowered when neighbouring  spins are
aligned. This effect is mostly due to the Pauli exclusion
principle. No such restriction applies if the spins are
anti-parallel. The geometrical structure of the model is shown in
Fig. \ref{fig-triangular}. Periodic boundary conditions were
employed during the simulations to reduce finite-size effects.

\setlength{\unitlength}{2mm}

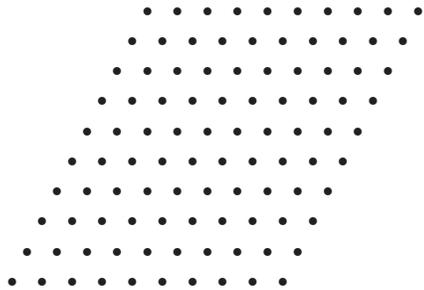
\begin{figure}[h]
  \centering
  \begin{picture}(20,20)
    \multiput(0,0)(1,2){10}{\circle*{0.5}}
    \multiput(2,0)(1,2){10}{\circle*{0.5}}
    \multiput(4,0)(1,2){10}{\circle*{0.5}}
    \multiput(6,0)(1,2){10}{\circle*{0.5}}
    \multiput(8,0)(1,2){10}{\circle*{0.5}}
    \multiput(10,0)(1,2){10}{\circle*{0.5}}
    \multiput(12,0)(1,2){10}{\circle*{0.5}}
    \multiput(14,0)(1,2){10}{\circle*{0.5}}
    \multiput(16,0)(1,2){10}{\circle*{0.5}}
    \multiput(18,0)(1,2){10}{\circle*{0.5}}
  \end{picture}
  \caption{Triangular Ising net.}
  \label{fig-triangular}
\end{figure}

Configuration ensembles were generated using both mixture of
Metropolis and single-cluster (multiple-cluster) methods on two
dimensional triangular lattice of size $L = 8 - 50$.  We outline the
procedure of single-cluster updating algorithm used to study the
Ising model on triangular lattice below:
\begin{enumerate}
  \item[(a)]
    Choose a random lattice site x as the first point of cluster c to built.
  \item[(b)]
    Visit all sites directly connecting to x, and add them to the cluster c with probability
    $P(S_x,S_y)=1-\exp\{min[0,2KS_xS_y]\}$, where y is the nearest neighbor of site x
    and $K$ is the inverse temperature.
  \item[(c)]
    Continue iteratively in the same way for the newly adjoined sites until the process stops.
  \item[(d)]
    Flip the all sites of the cluster c.
  \item[(e)]
    Repeat(a), (b), (c) and (d) sufficient times.
\end{enumerate}

We define a compound sweep as one single-cluster updating sweep
following with five Metropolis updating sweeps. In our simulations,
five compound sweeps were performed between measurements. We
observed that Metropolis -  single-cluster (multiple-cluster) and
compound sweep updating  techniques proved equally good for large
ensembles. However, in case of the compound sweep technique the
computational cost turned out to be far less. For this reason we
adapted this technique in our simulations.

We generated $1 \times 10^7$ ensembles at transition point for
each lattice size to performed the finite-size analysis. High
statistics could help us to obtain a more accurate result. Since
the ensembles were generated by Monte Carlo method, they were not
completely uncorrelated. The Jackknife procedure was employed
because it took the autocorrelation between the ensembles into
account and provided an improved estimate of the mean values and
errors.

\section{Results and discussion}

\begin{figure}[h]
  \centering
  \includegraphics[width=0.98\linewidth]{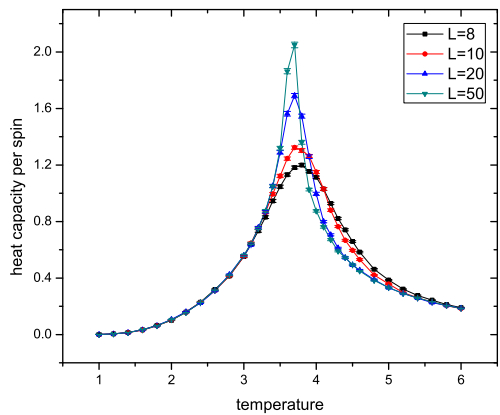}
  \caption{Plot showing the differing results of the
heat capacity with respect to temperature for varying lattice
size, $L \times L$.}
  \label{fig-capacity}
\end{figure}

From the fluctuation dissipation theorem, the heat capacity ($C$) is
the derivative of the energy with respect to temperature and has the
form
\begin{equation}
C=\frac{<E^2>-<E>^2}{L^2T^2}. \label{eqn2}
\end{equation}
For improved comparability among different system sizes, it is
better to compute the heat capacity per spin  which is displayed in
in Fig. \ref{fig-capacity}. Theoretical derivations suggest that the
heat capacity should behave like $\log\mid T-T_{c}\mid$ near the
critical temperature.  We notice a progressive steepening of the
peak as the lattice size increases, which illustrates a more
apparent phase transition. Note that due to finite size effects, the
peak is flattened and moved to the right. Our interest is to look
for such a transition point and investigate the critical phenomena
associated with the transition using finite-size scaling method
\cite{Fisher-1972-PRL,Landau-1976-PRB,Binder-1981-Z.Phys.B}

\subsection{Estimation of the transition temperatures}

\begin{figure}[h]
  \centering
  \includegraphics[width=0.98\linewidth]{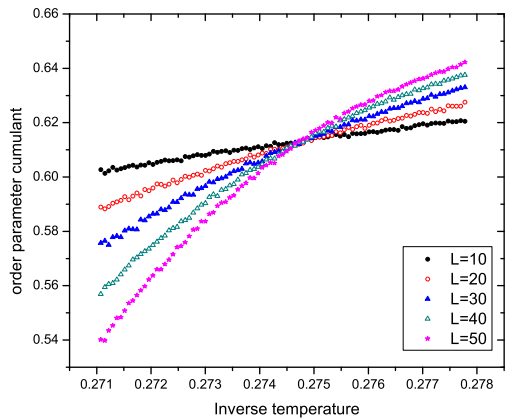}
  \caption{The fourth-order magnetic parameter cumulant plotted as a
         function of inverse temperature for several lattice sizes.}
  \label{fig-cum-mag}
\end{figure}

The fourth-order magnetic cumulant $U_L$, which is used to
estimate the transition temperature, is defined by the following
expression \cite{Binder-1981-Z.Phys.B}:
\begin{equation}
U(T,L)=1-\frac{<m^4(T,L)>}{3<m^2(T,L)>^2}, \label{eqn3}
\end{equation}
where $<m^k(T,L)>$ is the thermodynamic average value of the $k$th
power of the magnetic order parameter per spin for the lattice of
size $L$ with temperature $T$. The variation of $U(T,L)$, for a
given size $L$, with the temperature is illustrated in Fig.
\ref{fig-cum-mag}. To determinate the temperature $T_c(L,L')$, we
make use of the condition
\cite{Binder-1981-Z.Phys.B,Peczak-1991-PRB}
\begin{equation}
\left. \frac{U(T,L')}{U(T,L)} \right|_{T=T_c(L,L')}=1,
\end{equation}
where $L'$  and $L$ are two different lattice sizes. Thus we can
determine $T_{c}(L,L')$ by locating the interaction of these curves
- the so called ``cumulant crossing".

\begin{figure}[h]
  \centering
  \includegraphics[width=0.98\linewidth]{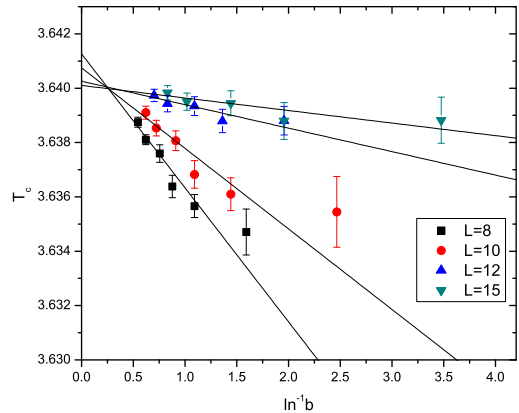}
  \caption{Estimates for $T_c$ plotted vs inverse logarithm of the scale
           factor $b=L'/L$ for serval lattice sizes.}
  \label{fig-transition-temperature}
\end{figure}

Fig. \ref{fig-transition-temperature}  shows the results for
$T_c(L,L')$ plotted as a function of the ratio $\ln ^{-1}(L'/L)$.
Keeping $L$ fixed, linear extrapolation is performed to obtain
$T_c(L,\infty)$ which corresponds to the critical temperature of
lattice size $L$. Results of the extrapolations for $8 \le L \le
15$ agree quite well within the errors. The transition temperature
for the infinite lattice is thus estimated as $T_c=3.6403(2)$,
which is much close to the exact value $4/\ln3$.

\subsection{Order of the transition}

\begin{figure}[h]
  \centering
  \includegraphics[width=0.98\linewidth]{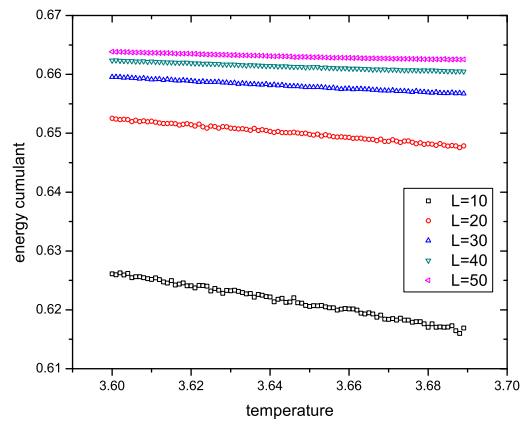}
  \caption{Plots of energy cumulant with respect to temperature for serval lattice sizes.}
  \label{fig-order-cum-eng}
\end{figure}

A better way to identify the order of the transition is the
internal-energy cumulant
\cite{Binder-1981-PRL,Binder-1984-PRB,Challa-1986-PRB,Ferrenberg-1988-PRL,Billoire-1990-PRB,Rastelli-2005-PRB}
defined by
\begin{equation}
V(T,L)=1-\frac{<E^4(T,L)>}{3<E^2(T,L)>^2}, \label{eqn4}
\end{equation}
where $<E^k(T,L)>$ is the thermodynamic average value of $k$th
power of internal energy for the lattice of size $L$ with
temperature $T$. $V(T,L)$ is a useful quantity since its behavior
at a continue phase transition is quite different from that at a
first order transition. This variable has a minimum (near the
critical point), and in the limit $L\rightarrow\infty$  achieves
the value $V^*=\frac{2}{3}$ for a continuous transition, whereas
$V^*<\frac{2}{3}$ is expected in the case of a first-order
transition. Plots of $V(T,L)$ versus temperature for several
lattice sizes are shown in Fig. \ref{fig-order-cum-eng}. The
graphs indicate that  the transition is not first-order since the
energy culumant $V(T,L)$ does not peak near the critical
temperature. This has also been observed in various other studies
 \cite{Challa-1986-PRB,Billoire-1990-PRB,Rastelli-2005-PRB}.

\begin{figure}[h]
  \centering
  \includegraphics[width=0.98\linewidth]{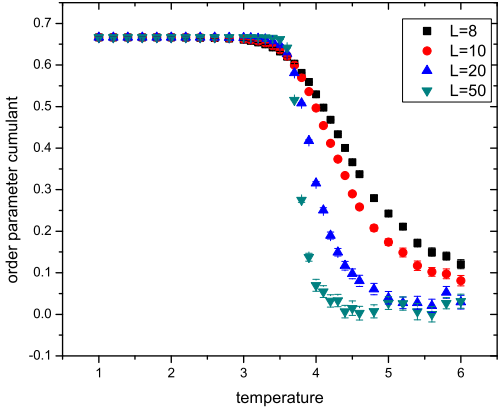}
  \caption{plots of order parameter cumulant with respect to temperature for serval lattice sizes.}
  \label{fig-order-cum-mag}
\end{figure}

Fig. \ref{fig-order-cum-mag} shows the order parameter cumulant
$U(T,L)$ as a function of temperature for serval lattice sizes. In
contrast with obtaining a negative minimum value in first order
transition, $U(T,L)$ drops from $2/3$ for $T<T_c$ to 0 for
$T>T_c$. This is in agreement with the continuous transition
\cite{Binder-1981-PRL}.

Since the magnetic susceptibility $\chi$ scales as $L^d$ for a
discontinuous transition and as $L^{\gamma/\nu}$ for a continuous
one \cite{Bunker-1993-PRB}, we can also determine the transition
order by $\chi$'s scaling behaviors. As discussed  below, we find
that $\chi$ scales as $L^{1.75223}$ for $T<T_c$ and as $L^{1.7555}$
for $T>T_c$, but not as $L^2$. Thus we can conclude that the
transition is continuous.

\subsection{Estimation of the critical exponents}

To get an estimation for the critical exponents, finite-size scaling
relations at critical point are used. For this purpose, we extract
the critical exponents, which are related to  specific heat, the
order parameter, and the susceptibility, from our simulation data at
$T_{c}(\infty )$ by using finite-size analysis
\cite{Fisher-1972-PRL,Landau-1976-PRB,Binder-1981-Z.Phys.B}.

\begin{figure}[h]
  \centering
  \includegraphics[width=0.98\linewidth]{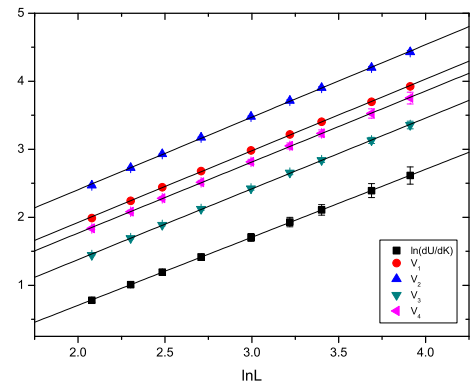}
  \caption{The critical exponent $\nu$ is determined from the slope
    of $\ln (dU_L/dK)$ and $V_i$ vs $\ln L$ at $T=3.6403$. Linear fittings
    result in $\nu=0.9995(21)$.}
  \label{fig-exp-nu}
\end{figure}

First we extract $\nu$ from $dU/dK$ which obey the following
relation
\begin{equation}
\frac{dU}{dK} \sim L^{1/\nu},
\end{equation}
Finite-difference derivative method can be used to determinate
$dU/dK$, but it is rather a poor choice  since the results are
highly  sensitively to the interval of the linear approximation.
To obtain a precise result of the critical exponent $\nu$, one can
write the derivative of $U$ with respect to $K$ as
\begin{eqnarray}
\frac{dU}{dK} &=&
\frac{1}{3<m^2>^2}\left[<m^4><E>-2\frac{<m^4><m^2E>}{<m^2>}
\right.
\nonumber\\
& & \left. +<m^4E>\right] \nonumber \\
  &=& <\frac{m^4(<E>+E)}{3<m^2>^2}>-<\frac{2m^2E<m^4>}{3<m^2>^3}>.
\end{eqnarray}
The plot of $\ln (dU/dK)$ versus $\ln L$ is shown in Figure
\ref{fig-exp-nu}.  The slope of the straight line obtained from the
log-log plot of the scaling relation corresponding to this quantity
gives the correlation length exponent $\nu=0.9995(21)$, which is
slightly different from the theoretical value. Since the critical
properties of $\ln (dU/dK)$ have a large lattice size dependence
\cite{Bunker-1993-PRB}, we could also extract the value of $\nu$
from other observables, which are defined as following
\cite{Ferrenberg-1991-PRB,Bunker-1993-PRB,Chen-1993-PRB}:
\begin{eqnarray}
V_1 &=& \ln \left[\frac{d\ln <m>}{dK}\right]= \ln \left[<E>-\frac{<mE>}{<m>}\right]\nonumber \\
    &=& \ln \langle \frac{E(<m>-m)}{<m>}\rangle, \\
V_2 &=& \ln \left[\frac{d\ln <m^2>}{dK}\right]=  \ln \langle
            \frac{E(<m^2>-m^2)}{<m^2>}\rangle,  \\
V_3 &=& 2[m]-[m^2],  \\
V_4 &=& 3[m^2]-2[m^3],
\end{eqnarray}
where
\begin{eqnarray}
[m^n] &=& \ln \frac{\partial<m^n>}{\partial K}
          \nonumber \\
      &=& \ln (<E><m^n>-<m^nE>)
          \nonumber \\
      &=& \ln <E(<m^n>-m^n)>.
\end{eqnarray}
The curves show a smooth behavior and linear fitting of the data is
adopted. We found the slopes of the straight lines of $V_i$, with
$i=1,2,3,4$, are very close to that of $\ln (dU/dK)$, as illustrated
in Fig. \ref{fig-exp-nu}.

\begin{figure}[h]
  \centering
  \includegraphics[width=0.98\linewidth]{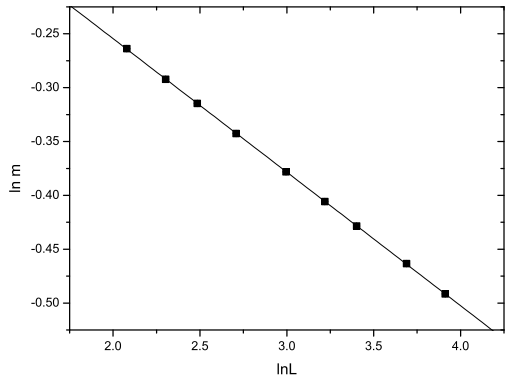}
  \caption{Plot of $\ln m$ as a function of $\ln L$ at $T=3.6403$.}
  \label{fig-exp-beta}
\end{figure}

Figure \ref{fig-exp-beta} displays the double-logarithmic plot of
the magnetization at transition point as a function of $L$.
According to the standard theory of finite-size scaling, at $T_c$
the magnetization per spin should obey the relation
\begin{equation}
m \sim L^{-\beta/\nu},
\end{equation}
for sufficiently large lattice size. The data falls nicely on a
linear curve and the slope of the curve gives an estimate for the
ratio of critical exponents $\beta/\nu=0.12400(18)$. Using our
estimate of $\nu=0.9995(21)$, we get the critical exponent of
magnetization $\beta=0.12394(32)$.

On finite lattices, the magnetic susceptibility per spin is given
by\cite{Binder-1969-Z.Phys.,Peczak-1991-PRB}
\begin{equation}
\chi=(L^2/T)<m^2>
\end{equation}
for $T>T_c$ and
\begin{equation}
\chi'=(L^2/T)(<m^2>-<m>^2)
\end{equation}
for $T<T_c$, and satisfy
\begin{equation}
\chi \sim L^{\gamma/\nu},\hspace{1.0cm} \chi' \sim
L^{\gamma'/\nu}.
\end{equation}

\begin{figure}[h]
  \centering
  \includegraphics[width=0.98\linewidth]{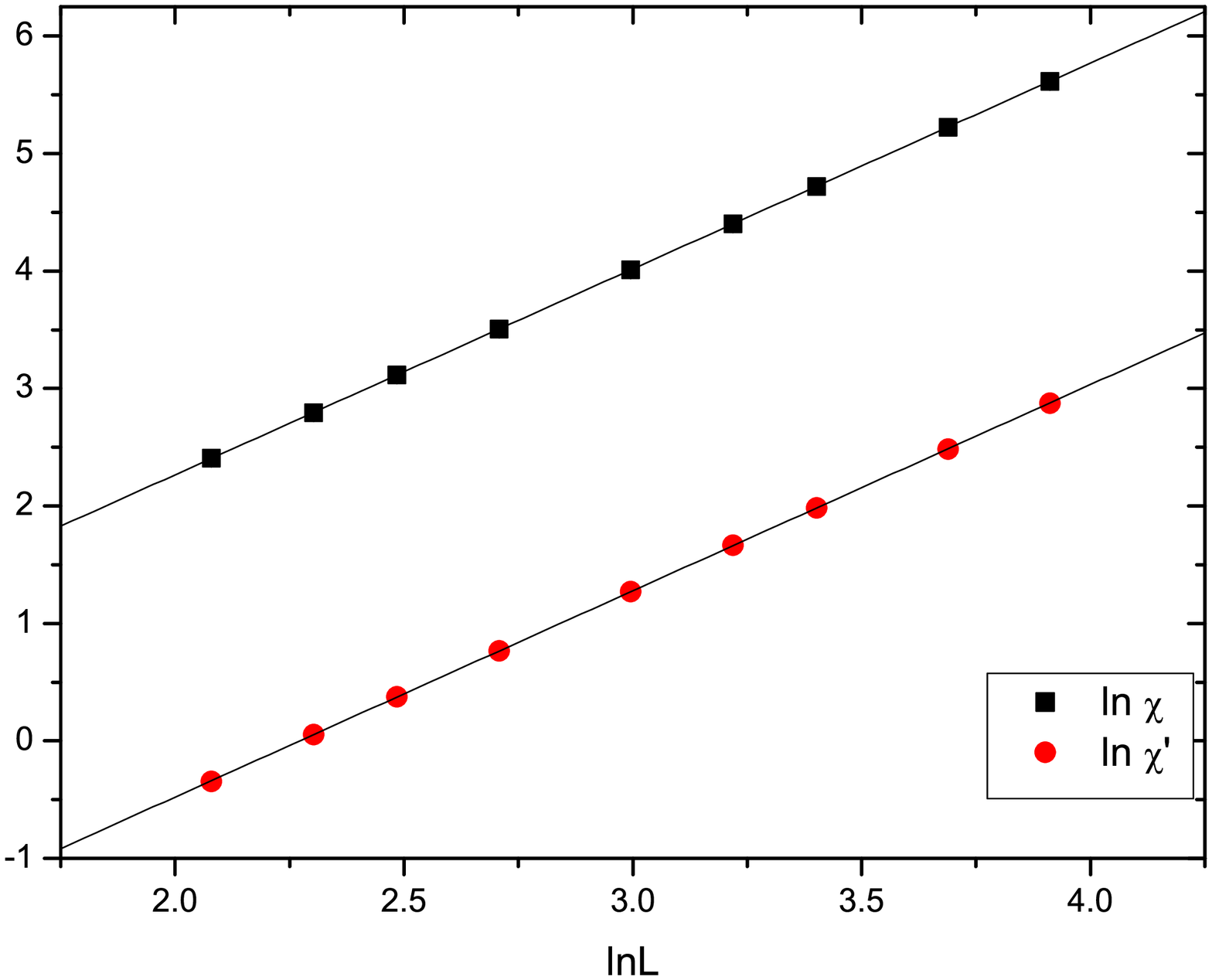}
  \caption{Plot of $\ln \chi$ and $\ln \chi'$ vs $\ln L$ at $T=3.6403$.}
  \label{fig-exp-gamma}
\end{figure}

As another estimation, we compute the magnetic susceptibility per
spin   by examining log-log plot shown in Fig. \ref{fig-exp-gamma}
of $\chi$ and $\chi'$ versus $L$ at the transition point. The data
displays a  smooth scaling behaviour. The slope of the linear fit to
the data gives estimates of $\gamma/\nu=1.75223(22)$ and
$\gamma'/\nu=1.7555(22)$, respectively. With our estimated values
$\nu=0.9995(21)$, we obtain $\gamma=1.7514(37)$ and
$\gamma'=1.7546(43)$, respectively. These results agree rather well
with the theoretical prediction.

With  three critical exponents $\nu$, $\gamma$, $\beta$
determined, the fourth exponent $\alpha$ is estimated using
scaling law
\begin{equation}
2\beta+\alpha+\gamma=2. \label{eqn17}
\end{equation}
This yields $\alpha/\nu=0.00077(420)$ and $\alpha=0.00077(420)$ for
$\gamma/\nu=1.75222(22)$, $\alpha/\nu=-0.0025(47)$ and
$\alpha=-0.0025(47)$ for $\gamma'/\nu=1.7555(22)$ respectively. We
can also estimate $\alpha$ from hyperscaling relation
\begin{equation}
d\nu+\alpha=2,
\end{equation}
where $d=2$ is dimension of the system. It gives a result
$\alpha=0.0010(42)$ and $\alpha/\nu=0.0010(42)$, which is consistent
with that given by scaling law.

Estimates of critical exponents from finite-size scaling at
critical temperature $T_c$ are summerized in Table \ref{tab-exp}.
\begin{table}[!h]
  \tabcolsep 2mm
  \caption{Estimates of critical exponents}
  \label{tab-exp}
\begin{center}
\begin{tabular}{c|c|c}
  \hline \hline
exponent        & value           &   exact value\cite{Fisher-1967-Rep.Prog.Phys.,Mon-1993-PRB} \\
  \hline
$\nu$           &   0.9995(21)            &   1   \\
$\beta/\nu$     &   0.12400(18)                     &   $1/8$   \\
$\gamma/\nu$    &   1.75223(22)($T>T_c$)            &   $7/4$   \\
$\gamma'/\nu$   &   1.7555(22)($T<T_c$)             &   $7/4$   \\
$\alpha/\nu$    &   0.00077(420)(by scaling)        &   0   \\
$\alpha/\nu$    &   0.0010(42)(by hyperscaling)     &   0   \\
  \hline \hline
\end{tabular}
\end{center}
\end{table}

To ensure the  validity of our estimate of $\alpha$, we compare it
with those obtained using the specific heat per spin from the
fluctuations of the total energy. For a continuous transition it
behaves as
\begin{equation}
C \simeq a+bL^{\alpha/\nu}. \label{eqn19}
\end{equation}
\begin{figure}[h]
  \centering
  \includegraphics[width=0.98\linewidth]{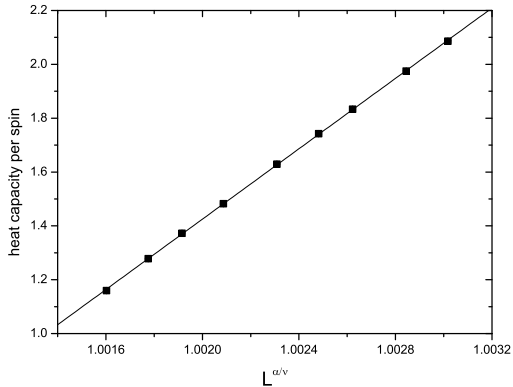}
  \caption{Plot of heat capacity $C$ with respect to $L^{\alpha/\nu}$,
      where $\alpha/\nu=0.00077$ and $C$ is evaluated at $T_c=3.6403$.}
  \label{fig-exp-alpha}
\end{figure}

Using $\alpha/\nu=0.00077(420)$ calculated above, the heat capacity
at critical temperature is plotted as a function of $L^{0.00077}$ in
Fig. \ref{fig-exp-alpha}. It was found that the evaluations of
(\ref{eqn17}) and (\ref{eqn19}) yield results very consistent within
statistical errors.  We find the values are in good agreement with
the theoretical ones. An over all error of about $0.1 - 0.8\%$ is
estimated for the values of the critical exponents

\section{Conclusions}

We have obtained  critical exponents for the Ising model with
nearest-neighbour interactions on a triangular lattice by  using a
mixture of single-cluster and Metropolis algorithm in our
simulations.  The data are analysed according to the finite-size
scaling theory.  The idea of using extensions of mixed algorithm to
estimate the critical exponents seems to supply a quite accurate
route for their estimation. In conclusion, it can be stated that the
compound update algorithm method with periodic boundary conditions
reproduces with a high accuracy the critical properties of the model
and confirms the prediction that first order transition is not
supported by the finite size behavior of the system. Our results
show that the several estimations for the critical exponents are in
good agreement with their theoretical values. We stress that the
present result for the exponent  is better than previous estimates
obtained by finite size scaling and Monte Carlo approach.

\section{Acknowledgements}

This work was supported in part by Guangdong Natural Science
Foundation(GDNSF) Grant No. 07300793. ML was supported in part by
the Guangdong Ministry of Education. We would like to express our
gratitude to the Theory Group at Sun Yat-Sen University for the
access to its computing facility.

\end{document}